\documentclass[a4paper,twocolumn,aps,pra]{revtex4}
\usepackage[utf8]{inputenc}
\usepackage{amsmath} 
\usepackage{mathtools}
\usepackage{amsfonts} 
\usepackage{amssymb}
\usepackage{graphics}
\usepackage{graphicx}
\usepackage{float}
\usepackage{physics}
\usepackage{color}
\usepackage{braket}
\usepackage{tikz}
\usepackage{pgfplots}
\usepackage{fancyhdr}
\usepackage{bbold}
\usepackage{hyperref}
\usepackage{listings}             
\usepackage{todonotes}
\usepackage{psfrag}

\usepackage{color}
\newcommand{\comment}[1]{}
\newcommand{\eq}[1]{\begin{equation}#1\end{equation}}
\newcommand{\ee}{\mathrm{e}}

\newcommand{\const}{\mathrm{const}}

\begin{document}

\title{Magnetization and entanglement after a geometric quench in the XXZ chain}

\author{Matthias Gruber}
\affiliation{%
  Institute of Theoretical and Computational Physics,
  Graz University of Technology, Petersgasse 16, 8010 Graz, Austria
}%

\author{Viktor Eisler}
\affiliation{%
  Institute of Theoretical and Computational Physics,
  Graz University of Technology, Petersgasse 16, 8010 Graz, Austria
}%

\begin{abstract}
We investigate the dynamics of the XXZ spin chain after a geometric quench, which is realized by connecting
two half-chains prepared in their ground states with zero and maximum magnetizations,
respectively. The profiles of magnetization after the subsequent time evolution are studied numerically by
density-matrix renormalization group methods, and a comparison to the predictions of generalized
hydrodynamics yields a very good agreement. We also calculate the profiles of entanglement entropy 
and propose an ansatz for the noninteracting XX case, based on arguments from conformal field theory.
In the general interacting case, the propagation of the entropy front is studied numerically both before and
after the reflection from the chain boundaries. Finally, our results for the magnetization fluctuations
indicate a leading order proportionality relation to the entanglement entropy.
\end{abstract}

\maketitle

\section{Introduction}

The dynamics of closed many-body quantum systems has been the
topic of a vast amount of research activities \cite{PSSV11,GE16},
with a special attention devoted to integrable models \cite{CEM16}.
The most commonly studied scenarios include the quantum quench \cite{EF16}
and the subsequent relaxation towards a stationary state \cite{VR16}.
A special feature of integrable systems is that, due to the existence
of stable quasiparticle excitations, a nonequilibrium steady state supporting
persistent currents may emerge.
The transport properties in systems driven by initial inhomogeneities have been the
subject of numerous investigations, see the review \cite{VM16} and references therein.

An important breakthrough in the understanding of transport
has been the development of a hydrodynamic theory \cite{BCDNF16,CADY16},
which could properly account for the nontrivial conservation laws of 
interacting integrable systems. Originally formulated for partitioned initial states
\cite{BCDNF16,CADY16,PDNCBF17}, the theory of generalized hydrodynamics
(GHD) has since been extended to include slowly varying inhomogeneities
\cite{DY17,DSY17,DDKY17,CDDKY17,BVKM17,BVKM18} and applied to a variety of
integrable models \cite{IDN17,DYC18,MVCRDL18,MBPC19}. Although GHD implies
in general a ballistic transport, recently it has been shown how diffusive mechanisms,
observed numerically in particular cases \cite{MMK17,LZP17,Stephan17,CDLV18},
can be incorporated into the theory \cite{DNBD18}

The GHD formalism has proved to be very successful in describing the time evolution
of magnetization or density profiles starting from an inhomogeneous inital state.
Another very important question is, however, how entanglement spreads out
in the time-evolved state. After a global quench in a homogeneous interacting system,
this can be answered by ascribing the growth of entropy to entangled pairs of quasiparticles
\cite{CC05}, and finding the corresponding entropy production rates \cite{AC17,AC18}.
Furthermore, since the description is based on quasiparticles, there have been attempts
to use the machinery of GHD to understand  entanglement evolution from inhomogeneous
initial states \cite{Alba18,BFPC18}.

The quasiparticle interpretation yields, by construction, a linear growth of entanglement
in time. In contrast, in a number of inhomogeneous situations such as front evolution
from domain walls or initial states created by slowly varying potentials
\cite{GKSS05,EIP09,SM13,EP14,DSVC17,DSC17,EB17,EME16,EM18},
it has been observed that the growth of entanglement is much slower, at most logarithmic.
In the quasiparticle picture, these correspond to zero entropy density states,
and the hydrodynamic information alone is not sufficient to account for the
growth of entanglement. For the particular case of the domain-wall quench,
further insight is provided by a special conformal field theory (CFT) treatment
\cite{DSVC17} which is, however, restricted to noninteracting fermions.
Hence the general mechanism leading to a sublinear growth of entanglement still
remains elusive.

In this paper we will study such a problem by considering the so-called geometric
quench \cite{MPC10,AHM14} in the XXZ spin chain or, equivalently,
fermions with nearest-neighbor interactions. The quench protocol simply consists of
releasing the ground state of a chain, by joining it to another one prepared in the vacuum.
We show that the magnetization profiles can be perfectly reproduced
using the GHD formalism and we find that they depend qualitatively on the sign of the
interaction, as opposed to the simpler setup of a domain-wall initial state \cite{CDLV18}.
Furthermore, with the help of some heuristic CFT arguments, we put forward an ansatz
for the entanglement profile in the noninteracting case, which gives a very good description
of the numerical data. Although the generalization to the interacting case does not seem
to be straightforward, our numerical results show a qualitatively similar behavior
of the entropy profiles, with expansion velocities that are obtained from the GHD solution
of the dynamics. Finally, by studying the fluctuations of the subsystem magnetization,
we observe that they are approximately proportional to the entropy within the entire front region,
which seems to hold for a large range of interactions.

The rest of the manuscript is structured as follows. In Section \ref{sec:model}
we introduce our setup and give a brief overview of the model. In Sec. \ref{sec:mag}
we present the magnetization profiles, starting with a review of GHD and followed
by our numerical results, whereas the entropy profiles are discussed in Sec. \ref{sec:ent}.
Reflections due to boundaries are studied in Sec. \ref{sec:refl} and a comparison between
magnetization fluctuations and entropy can be found  in Sec. \ref{sec:fluct}.
Our closing remarks are given in Sec. \ref{sec:conc} followed by an Appendix
containing details of the calculation for the edge profile.

\section{Model and setup\label{sec:model}}

We consider the XXZ spin chain which is given by the Hamiltonian
\begin{equation}
  H = J \sum_{j=1}^{L-1} \left( S^x_j S^x_{j+1} + S^y_j S^y_{j+1} + \Delta S^z_j S^z_{j+1} \right)
\label{hxxz}
\end{equation}
where $S^{\alpha}_j$ are spin-1/2 operators acting on site $j$, $J$ is the coupling and
$\Delta$ the anisotropy parameter. We set $J=1$ and consider open boundary conditions
on a chain of length $L$.
The XXZ model is equivalent to a chain of spinless fermions with nearest-neighbor interactions of
strength $\Delta$, with $\Delta=0$ corresponding to the free-fermion point.

\begin{figure}[H]
  \centering
  \includegraphics[]{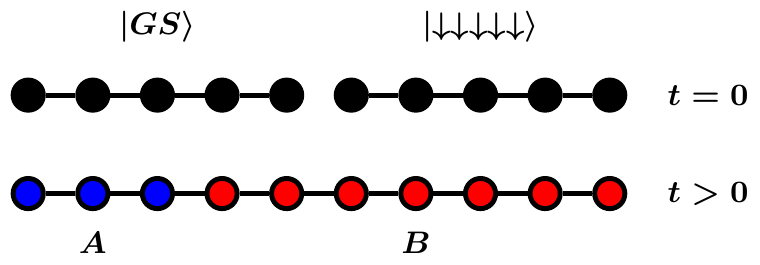}
  \caption{Setup of the geometric quench.}
  \label{fig:setup}
\end{figure}

The protocol of the geometric quench is illustrated in Fig. \ref{fig:setup}.
Initially, the chain is split in two halves and the left hand side is initialized in
the ground state $\ket{GS}$ of an XXZ chain of length $L/2$. On the other hand, the right
half-chain is prepared in the fully polarized state $\ket{\downarrow\downarrow\downarrow\dots}$,
or the vacuum state in the fermionic language. Subsequently, the two half-chains are joined together
at $t = 0$ and the system is let evolve unitarily 
\eq{
\ket{\psi(t)}= 
\ee^{-iHt} \ket{GS} \otimes \ket{\downarrow\downarrow\downarrow\dots}
\label{psit}}
governed by the Hamiltonian in Eq. \eqref{hxxz}.
In other words, we would like to study how the ground state prepared on a half-chain
expands into vacuum after an instantaneous change of geometry (i.e. the size of the chain),
hence the term geometric quench. We are primarily interested in the magnetization
$\braket{S^z_j}\equiv \bra{\psi(t)}S^z_j\ket{\psi(t)}$ and the entanglement profile,
as measured by the entanglement entropy between two segments $A$ and $B$,
as depicted in Fig. \ref{fig:setup}.

The ground state of the XXZ chain can be constructed with the help of Bethe Ansatz \cite{Takahashi,Franchini}.
Here we will focus on the regime $|\Delta| < 1$ where the ground state is a gapless
Luttinger liquid \cite{Giamarchi}, and we use the standard parametrization $\Delta = \cos(\gamma)$.
The quasiparticle excitations of the XXZ chain are created upon the vacuum state
$\ket{\downarrow\downarrow\downarrow\dots}$ and are labelled by their rapidity $\lambda$.
They satisfy appropriate quantization conditions, as given by the roots of the Bethe equations.
In particular, the ground state involves only magnons with real $\lambda$,
but in general the solutions admit a family of string excitations \cite{Takahashi},
corresponding to roots parallel to the imaginary axis.
In the thermodynamic limit $L \to \infty$, and in the zero-magnetization sector, the roots on the real axis become continuous 
and their density $\rho(\lambda)$ satisfies the linear integral equation
\eq{
\rho(\lambda) +
\int_{-\infty}^{\infty} 
\frac{\dd \mu}{2\pi} \mathcal{K}(\lambda-\mu) \rho(\mu)=
\frac{ p'(\lambda)}{2\pi} .
\label{rho}}
Note that the r.h.s. of Eq. \eqref{rho} contains the derivative of the bare momentum
$p'(\lambda) = \theta'_1(\lambda)$,
while the integral on the left is due to elastic scattering between quasiparticles,
with the kernel $\mathcal{K}(\lambda)= \theta'_2(\lambda)$ being the differential scattering phase.
Both of them are given via
\begin{equation}
  \theta'_n(\lambda) = \frac{\sin(n \gamma)}{\cosh(\lambda)-\cos(n \gamma)},
  \qquad n=1,2 \, .
  \label{thetan}
\end{equation}

In fact, Eq. \eqref{rho} is just a simple example of the so-called dressing operation,
where a certain function of the rapidity gets modified by
the presence of the other quasiparticles.
The dressed version $f^{dr}$ of a bare function $f$ is defined as
the solution of
\begin{equation}
  f^{dr}(\lambda) + \int_{-\infty}^{\infty} \frac{\dd \mu}{2 \pi}
  \mathcal{K}(\lambda - \mu) n(\mu) f^{dr}(\mu) = f(\lambda) \, ,
  \label{eq: dress}
\end{equation}
which is a Fredholm-type integral equation and can be solved numerically \cite{atkinson2008fredholm}.
Here, $n(\mu)$ is the occupation function, i.e. the ratio of the particle density (occupied rapidities)
and the total root density, including the density of holes (unoccupied rapidities).
However, since the XXZ ground state does not contain holes, one has $n(\mu) \equiv 1$.
Hence, the root density is just proportional to the derivative of the dressed momentum,
$2\pi \rho(\lambda) = p'^{dr}(\lambda)$. Another important quantity we shall need
is the dressed quasiparticle energy $e^{dr}(\lambda)$, which follows from \eqref{eq: dress}
with the bare energy given by
\begin{equation}
  e(\lambda) = \frac{-\sin^2(\gamma)}{\cosh(\lambda) -\cos(\gamma)} \, .
  \label{elam}
\end{equation}

On the numerical side, we carry out density-matrix renormalization group (DMRG)
calculations \cite{Schollwoeck11}, using the ITensor C++ library \cite{itensor}.
The ground-state search is performed by applying DMRG on the left half-chain,
whereas the vacuum state on the right half-chain has a trivial matrix product state representation.
After the quench, the time evolution is done by applying tDMRG with a timestep of $\delta t = 0.05$,
a truncated weight of $10^{-10}$ and a maximum bond dimension of $\chi_{max} = 1200$.

\section{Magnetization Profiles\label{sec:mag}}

We start our study of the geometric quench with a discussion of the magnetization profiles.
Before presenting our numerical results, we shall introduce an efficient method that has been
developed recently for the study of transport in integrable systems.

\subsection{Generalized hydrodynamics}

The understanding of time evolution in integrable models due to initial inhomogeneities
has recently come to a breakthrough by the development of generalized hydrodynamics
\cite{BCDNF16, CADY16}.
The idea of GHD is to give an effective description of the dynamics and the underlying
state at a hydrodynamic scale. Indeed, in interacting integrable models the quasiparticle excitations
are moving freely, experiencing only phase shifts due to the scattering on other quasiparticles.
One then assumes that, for large times $t$ and large distances $x$ from the inhomogeneity, 
a dynamical equilibrium emerges, and the system is described by a local quasi-stationary state (LQSS).

For the class of initial states, where the inhomogeneity is solely due
to the junction of two, otherwise homogeneous states without any inherent length scale,
the LQSS depends only on the ray variable $\zeta = \frac{x}{t}$.
Assuming that there is only one type of quasiparticles involved (such as for the geometric quench),
specifying the LQSS amounts to finding the particle density $\rho_{\zeta}(\lambda)$
that varies along the rays. The kinetic theory of quasiparticles eventually leads to the continuity equation
\cite{BCDNF16, CADY16}
\begin{equation}
  \partial_t \rho_{\zeta}(\lambda) + \partial_x(v(\lambda) \rho_{\zeta}(\lambda)) = 0 \, ,
  \label{GHD}
\end{equation}
where the velocity $v(\lambda)$ is given by the dressed quasiparticle group velocity
\begin{equation}
  v(\lambda) = \frac{e'^{dr}(\lambda)}{p'^{dr}(\lambda)}.
  \label{vdr}
\end{equation}

The GHD equation \eqref{GHD} could be interpreted as an infinite set
of continuity equations for each $\lambda$, corresponding to the infinite set of
conserved charges that are present for integrable models.
Despite its apparent simplicity, one should stress that the solution of \eqref{GHD} is,
in general, nontrivial since the dressed velocity \eqref{vdr} itself depends on the
quasiparticle density. Indeed, the dressing operation \eqref{eq: dress}
contains information about the full occupation function $n_\zeta(\lambda)$, and thus
Eq. \eqref{GHD} has to be solved self-consistently. However, if the densities depend
only on the ray variable $\zeta$,  the GHD equation could be shown to simplify to
\cite{BCDNF16, CADY16}
\eq{
(\zeta - v(\lambda)) \, \partial_\zeta n_\zeta(\lambda) = 0 \, ,
\label{GHDn}}
which has the piecewise continuous solution
\begin{equation}
  n_{\zeta}(\lambda) = 
  \Theta(v(\lambda) - \zeta) n_L(\lambda)  + 
  \Theta(\zeta - v(\lambda)) n_R(\lambda) \, ,
  \label{nzeta}
\end{equation}
where $\Theta$ is the Heaviside step function and $n_{L/R}(\lambda)$
is the initial occupation on the left/right half-chain.

The solution \eqref{nzeta} has a clear physical interpretation, namely that
the information on the initial occupations gets transported by the quasiparticles.
Along a given ray $\zeta>0$ in the r.h.s. of the chain, only the quasiparticles
emitted from the left half-chain become visible that have sufficient velocity
$v(\lambda) > \zeta$ to arrive there. Similarly, for $\zeta<0$ the quasiparticles
are emitted from the right half-chain and propagate to the left.
Thus, for the simple initial states considered here, solving the
GHD equation boils down to determining the solution to $v(\lambda) = \zeta$,
where the dressing of the velocity is calculated with respect to the occupation function
in \eqref{nzeta}.

\begin{figure}[H]
  \centering
  \includegraphics[]{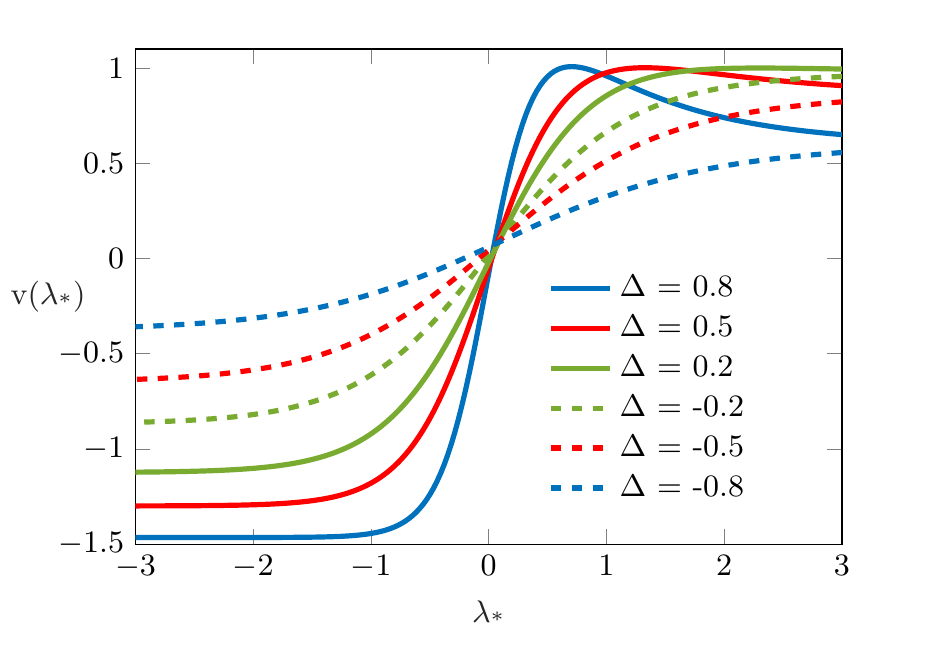}
  \caption{Dressed velocity $v(\lambda_*)$ corresponding to the occupation function
  in Eq. \eqref{nzeta1}, for several values of $\Delta$.}
      \label{fig:vls}
\end{figure}

The situation further simplifies for the geometric quench, since the
ground-state occupation is given by $n_L(\lambda) = 1$, whereas for
the vacuum one trivially has $n_R(\lambda) = 0$.
We shall first assume, that the dressed velocity is a monotonically increasing
function with a unique solution $v(\lambda_*) = \zeta$ for each $\zeta$ and hence
\eq{
n_{\zeta}(\lambda) = \Theta(\lambda-\lambda_*) \, .
\label{nzeta1}}
One has thus the condition that the function
\begin{equation}
  v(\lambda_*) = \frac{e'^{dr}(\lambda_*)}{p'^{dr}(\lambda_*)}
  \label{vls}
\end{equation}
has to be monotonously increasing, when the dressing is evaluated with the occupation
in \eqref{nzeta1}, i.e. the integrals in \eqref{eq: dress} are carried out over
$\left[\lambda_*,\infty\right)$. The velocity \eqref{vls} can be evaluated numerically
and the result is shown in Fig. \ref{fig:vls} for various $\Delta$.
One can see clearly, that our assumption is satisfied only for attractive interactions
$\Delta<0$, whereas for the repulsive case $\Delta>0$ the velocity $v(\lambda_*)$
develops a maximum.

The above discrepancy can be understood as follows. For $\Delta<0$, the maximum
velocity occurs for $\lambda_* \to \infty$, which gives the expansion velocity of the
front into vacuum. Note that in this limit the occupation \eqref{nzeta1} vanishes completely,
and thus the group velocity is given by its bare (undressed) value
\eq{
v_0(\lambda) = \frac{e'(\lambda)}{p'(\lambda)}=
\frac{\sin(\gamma)\sinh(\lambda)}{\cosh(\lambda)-\cos(\gamma)} \, .
\label{v0}}
In particular, one has
\eq{
v_0(\lambda\to\infty) = \sin (\gamma) \, ,
}
which turns out to be the real maximum for $\Delta<0$. However, for $\Delta>0$,
the equation $v'_0(\tilde\lambda)=0$ has a nontrivial solution with
\eq{
\cosh \small( \tilde \lambda \small) = \frac{1}{\Delta} \, .
\label{lamt}}
The maximum velocity thus occurs at a finite value of the rapidity,
and one obtains $v_0(\tilde\lambda)=1$, independently of $\Delta$.
Consequently, the ansatz for the occupation function has to be
modified as
\eq{
n_{\zeta}(\lambda) = 
\Theta(\lambda - \lambda_1) \Theta(\lambda_2 - \lambda) \, ,
\label{nzeta2}}
where the velocities must satisfy
\begin{equation}
  v(\lambda_1) = v(\lambda_2) = \zeta \, .
  \label{vl12}
\end{equation}
Note that the rapidities $\lambda_1<\tilde\lambda < \lambda_2$ are located on
different sides of the maximum and can be found iteratively.

Interestingly, the GHD solution for the geometric quench yields different
vacuum expansion velocities, with the rightmost ray given by
$\zeta_{max}=1$ and $\zeta_{max}=\sin (\gamma)$ for positive and negative
values of $\Delta$, respectively. Note, however, that by decreasing $\zeta$,
the solution $\lambda_2$ of \eqref{vl12} eventually goes to infinity,
and thus the ansatz \eqref{nzeta2} actually goes over to \eqref{nzeta1}
with $\lambda_1 \to \lambda_*$. In particular, one finds that the minimum
of the dressed velocity occurs for $\lambda_* \to -\infty$ (see Fig. \ref{fig:vls}),
where \eqref{nzeta1} simply corresponds to the ground-state occupation.
Therefore, the leftmost ray is given via the spinon velocity \cite{Franchini}
\eq{
\zeta_{min} = - v_s = -\frac{\pi}{2}\frac{\sin(\gamma)}{\gamma} \, .
}

Finally, in order to obtain the magnetization profile, one needs
the particle density $\rho_{\zeta}(\lambda)$. This is given explicitly by
\eq{
\rho_\zeta(\lambda) = n_{\zeta}(\lambda) \frac{ p'^{dr}(\lambda)}{2\pi} \, ,
\label{rhoz}}
where the dressing is calculated with an occupation $n_{\zeta}(\lambda)$ that
corresponds to either \eqref{nzeta1} or \eqref{nzeta2}.
In turn, the magnetization is given by
\begin{equation}
  \expval{S^z} = -\frac{1}{2} + \int \rho_\zeta(\lambda) d\lambda \, .
  \label{szghd}
\end{equation}

Although in general the GHD ansatz requires a numerical solution of the
integral equations for the dressing, there is one particular regime where
an approximate analytical result can be given. Namely, for $\Delta>0$
the magnetization profile around the $\zeta_{max} = 1$ edge can be 
obtained to leading order via a perturbative solution, with the details
of the calculation presented in the Appendix.
Indeed, the edge regime $\zeta \to 1$ corresponds to occupied
rapidities in the interval $\left[\lambda_1,\lambda_2\right]$, where
$\lambda_{1,2} = \tilde{\lambda} \mp \epsilon$ and we assume $\epsilon \ll 1$.
The perturbative solution of Eq. \eqref{vl12} then gives to lowest order
\begin{equation}
  \epsilon(\zeta) = \sqrt{2 (1-\zeta)} \tan(\gamma) \, .
  \label{epsz}
\end{equation}
Moreover, the profile can also be approximated by noting that the
integral in \eqref{szghd} is taken over a very short interval
around $\tilde \lambda$ and the effect of dressing in \eqref{rhoz}
can be neglected. This yields
\begin{equation}
  \expval{S^z} \approx - \frac{1}{2} + \frac{1}{\pi} p'(\tilde{\lambda}) \epsilon(\zeta) = \\
			   	  - \frac{1}{2} + \frac{1}{\pi}\sqrt{2(1-\zeta)}
  \label{szedge}
\end{equation}
and thus one has a leading square-root singularity of the edge profile,
which is independent of $\Delta$. Interestingly, the very same behavior
was found for the edge profile in the XXZ chain with a magnetic field gradient
\cite{EB17}.

To conclude this section, one should remark that the analytical form of the
entire profile can be found explicitly for the noninteracting XX chain \cite{AHM14}.
There, instead of rapidities, one can simply work with
momentum modes, and the velocities $v(q)=\sin q$ are given by the derivative of
the dispersion, independently of the occupation function. The magnetization
along the ray $\zeta$ then follows from the number of modes that satisfy
$v(q)>\zeta$. In general, $v(q)=\zeta$ has two solutions
for $|\zeta|<1$, given by $q_\pm = \pi/2 \pm \arccos \zeta$. Note, however, that the initial state
on the l.h.s. is the half-filled ground state and thus $|q|\le\pi/2$ must be satisfied.
Hence, the modes that contribute lie in the interval $\left[q_-,\pi/2\right]$ and the
magnetization reads
\eq{
\braket{S^z} = -\frac{1}{2} + \mathcal{N}(\zeta), \quad
\mathcal{N}(\zeta) = \frac{1}{2\pi} \arccos \zeta \, .
\label{szxx}}
%

\begin{figure*}[t]
  \centering
  \includegraphics[trim={ 15px 35px 90px 20px}]{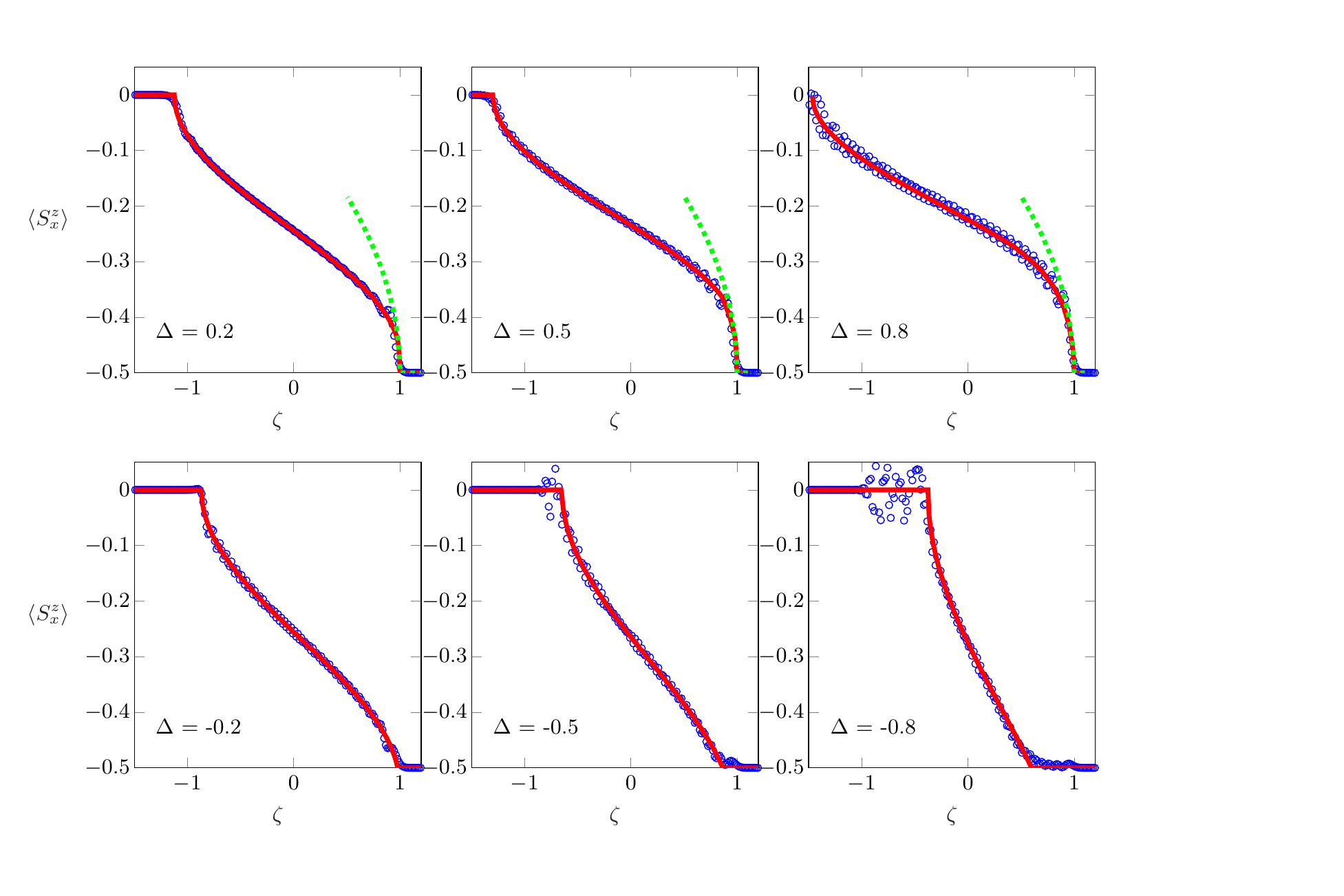} 
  \caption{Magnetization profiles $\expval{S^z_x}$ after the geometric quench as a function of $\zeta = x/t$,
   for various values of $\Delta$. The symbols (blue) are the results of DMRG calculations, whereas the
   solid lines (red) are obtained from the numerical solution of the GHD ansatz. For $\Delta>0$ the
   approximation near the edge, Eq.~\eqref{szedge}, is shown by dashed lines (green).}
      \label{fig:sz}
\end{figure*}

\subsection{Numerical results}

We now present our numerical results from DMRG calculations
and compare them to the $\expval{S^z}$ profiles as obtained from \eqref{szghd}
by solving the GHD equations. In Fig. \ref{fig:sz} the magnetization profiles are
reported for a system with $L = 200$ sites and for a fixed time $t = 64$ after the quench.
Instead of the lattice site $j=1,\dots,L$, we introduce the (half-integer) distance $x=j-(L+1)/2$
from the junction of the half-chains to index the sites, and plot the data $\braket{S^z_x}$
against the ray variable $\zeta=x/t$.
For all the anisotropies presented, one generally observes a very good agreement
between the DMRG data and the GHD solution. There are, however, some extra
features that should be discussed.

First, for $\Delta>0$, the right edge of the front indeed lies at $\zeta_{max}=1$, 
as predicted by GHD and the ansatz \eqref{nzeta2} for the occupation provides,
up to oscillations, a very good description of the edge regime.
However, although the approximate solution in Eq. \eqref{szedge},
shown by the green dashed lines in Fig. \ref{fig:sz}, seems to capture the leading
behavior of the edge, its applicability is restricted to a rather small neighborhood
of $\zeta=1$. As further discussed in the Appendix, this is due to the fact that the
solution \eqref{epsz} which gives the interval of occupied rapidities actually fails to
satisfy $\epsilon \ll 1$, unless $1-\zeta$ is chosen to be extremely small.
In particular, $\epsilon$ diverges for $\Delta=0$ and the approximation improves
as $\Delta \to 1$.

On the other hand, for $\Delta<0$, the GHD edge is given by $\zeta_{max}=\sin (\gamma)$, 
whereas the density can be clearly seen to extend beyond this value up to $\zeta \approx 1$.
Moreover, the GHD profile shows a qualitatively different behavior around $\zeta_{max}$,
where the square-root singularity seems to be replaced with a linear profile.
In fact, this is very reminiscent to the case of the domain-wall initial state
$\ket{\uparrow\uparrow\uparrow\dots} \otimes \ket{\downarrow\downarrow\downarrow\dots}$,
where the analytical GHD profile can be obtained \cite{CDLV18} and the edge behavior
has recently been investigated in detail \cite{BK18,Stephan19}.
In particular, the tail has been interpreted as a dilute regime of quasiparticles,
where the interactions renormalize to zero and the edge $\zeta=1$
corresponds to the free magnon velocity \cite{Stephan19}.

\begin{figure}[H]
  \centering
  \noindent
  \includegraphics[]{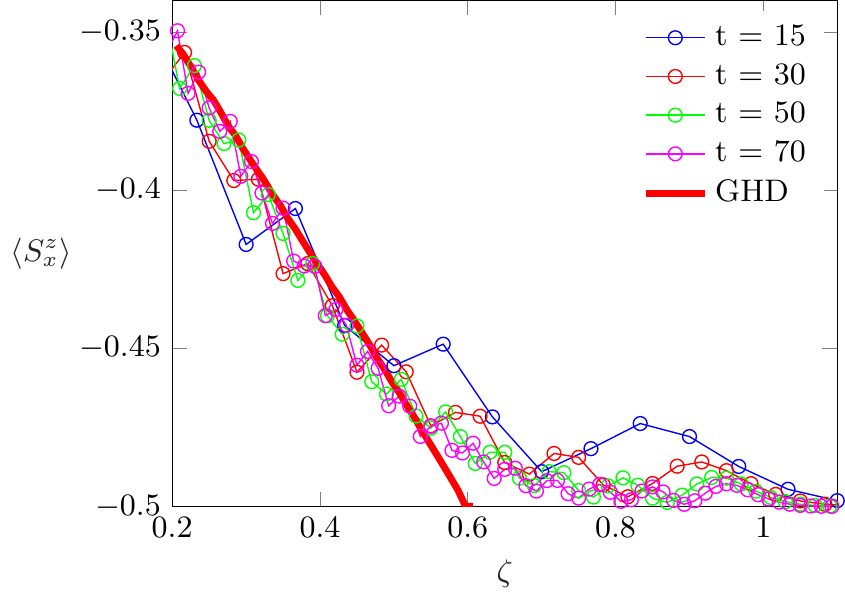}
  \caption{Edge profiles for $\Delta = -0.8$ and various times $t$, plotted against $\zeta=x/t$.
  The GHD solution is shown by the red solid line.}
      \label{fig:szedge}
\end{figure}

To have a better overview of the situation for the geometric quench, we show in
Fig. \ref{fig:szedge} the magnified edge region for $\Delta=-0.8$ and various times $t$.
One can see a slow decrease of the scaled profiles in the regime
$\zeta_{max}<\zeta<1$, suggesting that the tail should indeed contain only a
finite number of particles that could escape from the bulk of the front region.
We expect that, when plotted against $\zeta$, the tail should vanish in the $t\to\infty$
limit, as the escaped density becomes smeared out in an infinitely large region.
Note, however, that the results of Ref. \cite{Stephan19} for the domain-wall quench
are also compatible with a logarithmic increase in time of the overall number of
particles in the tail regime. A detailed analysis of the tail would require
much more numerical effort and is beyond the scope of the present manuscript.

One should also comment on the left edge of the profile, i.e. the front region that connects
to the ground state outside the light cone. As pointed out before, the GHD ansatz suggests that
the left edge should extend with the spinon velocity, i.e. the speed of the excitations above
the zero-magnetization background.
This seems to be in perfect accordance with the numerical data for $\Delta>0$.
In the attractive ($\Delta<0$) case, however, one observes very strong oscillations
beyond the GHD edge $\zeta < \zeta_{min}$. We believe that, similarly to the right edge,
this feature is due to a small number of particles that escape from the attractive bulk
of the front. Note also, that the GHD edge seems to have a square-root behavior for
all values of $\Delta$. However, a perturbative treatment is more complicated in this case,
since one has to consider the perturbation around the completely filled ground state,
instead of the vacuum.

Finally, it is interesting to note that in the limit
$\Delta \to -1$ one has $\zeta_{min}=\zeta_{max}=0$,
and thus the bulk of the front region vanishes completely.
This is a clear signal of subballistic transport in
the regime $\Delta \le -1$. On the other hand, the
limit $\Delta \to 1$ shows no singular features,
suggesting that the $\Delta>1$ regime is smoothly
connected and the ballistic nature of the dynamics
is preserved. These expectations seem to be
confirmed by our DMRG numerics.

\section{Entropy Profiles\label{sec:ent}}

The front dynamics can be further characterized by calculating the entanglement
entropy $S=-\Tr \rho_A \ln \rho_A$ for a given bipartition of the chain (see Fig. \ref{fig:setup})
where $\rho_A$ is the corresponding reduced density matrix. The entanglement profile
is obtained at a fixed time $t$ by varying the boundary between the subsystems
$A=\left[1,L/2+r\right]$ and $B=\left[L/2+r+1,L\right]$. In particular, $r=0$ corresponds to a
bipartition across the initial cut, the case which was already considered in Ref. \cite{AHM14}.

As opposed to the magnetization, the entanglement profile is more complicated to be captured
within the hydrodynamic approach. Indeed, although there has been much progress in
understanding the entropy evolution in terms of the quasiparticle picture \cite{AC17,AC18},
these results are restricted to quench scenarios where the growth is linear in time.
In contrast, it has already been observed in \cite{AHM14} that the geometric quench
induces a logarithmic entropy growth for $r=0$, which is also a characteristic of local quench
protocols \cite{EP07,EKPP08,CC07,SD11}. 

We first consider the noninteracting ($\Delta=0$) case where, invoking results from
CFT and with some heuristic arguments, we are able to provide an ansatz
for the full entanglement profile.

\subsection{XX chain}

To find a quantitative description of the entropy profile, there are some features to be
noted about the structure of the hydrodynamic state described above Eq. \eqref{szxx}.
First, the fermionic density $\mathcal{N}(\zeta)$ is exactly one-half of the corresponding
one for a domain-wall initial state \cite{ARRS99,AKR08}, where the occupied modes
$\left[q_-,q_+\right]$ are not restricted below the Fermi-level $q_F=\pi/2$. Hence,
the LQSS after the geometric quench is reminiscent to that of the domain-wall problem,
but differs by the presence of a sharp Fermi-edge. We thus argue that the entropy can be
obtained as a sum of two contributions, due to the spatially varying occupation
and to the Fermi-edge singularity, respectively.

The contribution from the Fermi edge can be identified by recalling the results for the
local quench, where two half-filled semi-infinite chains are joined together \cite{CC07}.
Indeed, since the initial filling is unbiased, the time evolution is entirely due to
the presence of two Fermi edges at momenta $q_F=\pm \pi/2$. The resulting entropy
profile can be obtained via CFT \cite{CC07} and reads
\eq{
S_{loc}=
\frac{1}{6}\ln (t^2-r^2) \, ,
\label{sloc}}
where we have ignored the nonuniversal constant which is independent
of both $t$ and $r$. This result can also be generalized to finite-size chains
by substituting the corresponding chord-variables \cite{SD11}
\eq{
t \to \frac{L}{\pi} \sin\left(\frac{\pi t}{L}\right), \quad
r \to \frac{L}{\pi} \sin\left(\frac{\pi r}{L}\right).
\label{chord}}

It is important to stress that the result \eqref{sloc} and \eqref{chord}
gives the entropy profile resulting from two Fermi edges, whereas
we need only the contribution from $q_F=\pi/2$, i.e. from the right-moving
wavefront. Thus, using trigonometric identities we rewrite
\eq{
S_{loc}= \frac{1}{6}\ln \left[
 \frac{L}{\pi} \sin\left(\frac{\pi (t-r)}{L}\right)
 \frac{L}{\pi} \sin\left(\frac{\pi (t+r)}{L}\right)\right],
\label{slocfin}}
which has exactly the desired additive form, with the arguments $t \mp r$
corresponding to the Fermi edges $q_F = \pm \pi/2$, respectively.

The second piece of contribution we have to identify is due to the
space-dependent occupation. As we have already remarked, this
should be closely related to the domain-wall problem, where the
entropy profile is also known explicitly \cite{EP14}. In fact, the solution
can be found via a curved-space CFT approach \cite{DSVC17}, by identifying
the underlying curved metric \cite{ADSV16} and mapping it
conformally onto a flat one on the upper half plane. The result can
be cast in the form
\eq{
S_{dw} = \frac{1}{6} \ln (\mathcal{L} \sin q_F(r/t)),
\label{sdw}}
where the conformal length is given by
\eq{
\mathcal{L} = 
t \left[1-\left(\frac{r}{t} \right)^2\right] .
\label{l}}
Note that \eqref{sdw} contains a nonuniversal part with
$\sin q_F(r/t)$ being the spatially varying Fermi velocity, 
where $q_F(x)=\arccos(x)$. This term plays the
role of a cutoff renormalization in the CFT picture.

We now give a heuristic argument on how to modify the expression
in \eqref{sdw} in order to get the result for the geometric quench.
As already pointed out, the fermionic density for the geometric quench
is exactly the half of that in the domain-wall case, by restricting to the modes
with $q \in \left[q_-,\pi/2\right]$. Due to the particle-hole symmetry of the problem,
one could also have worked with the modes $q \in \left[\pi/2,q_+\right]$ and
arrive to the same result. Thus, assuming that the universal entropy contribution
of the domain-wall problem could, in some way, be written as a sum over modes,
this symmetry argument implies that the universal contribution
to the geometric quench should be $\frac{1}{12}\ln \mathcal{L}$.
Moreover, one should also take into account the halved density
when considering the nonuniversal piece, where for the geometric
quench one has $q_F(x)=\pi \mathcal{N}(x) =\arccos(x)/2$, such that
\eq{
\frac{1}{6} \ln (\sin q_F(r/t)) = 
\frac{1}{12} \left[\ln \left(1-\frac{r}{t}\right) - \ln 2 \right] .
}

\begin{figure}[htb]
\center
\includegraphics[width=\columnwidth]{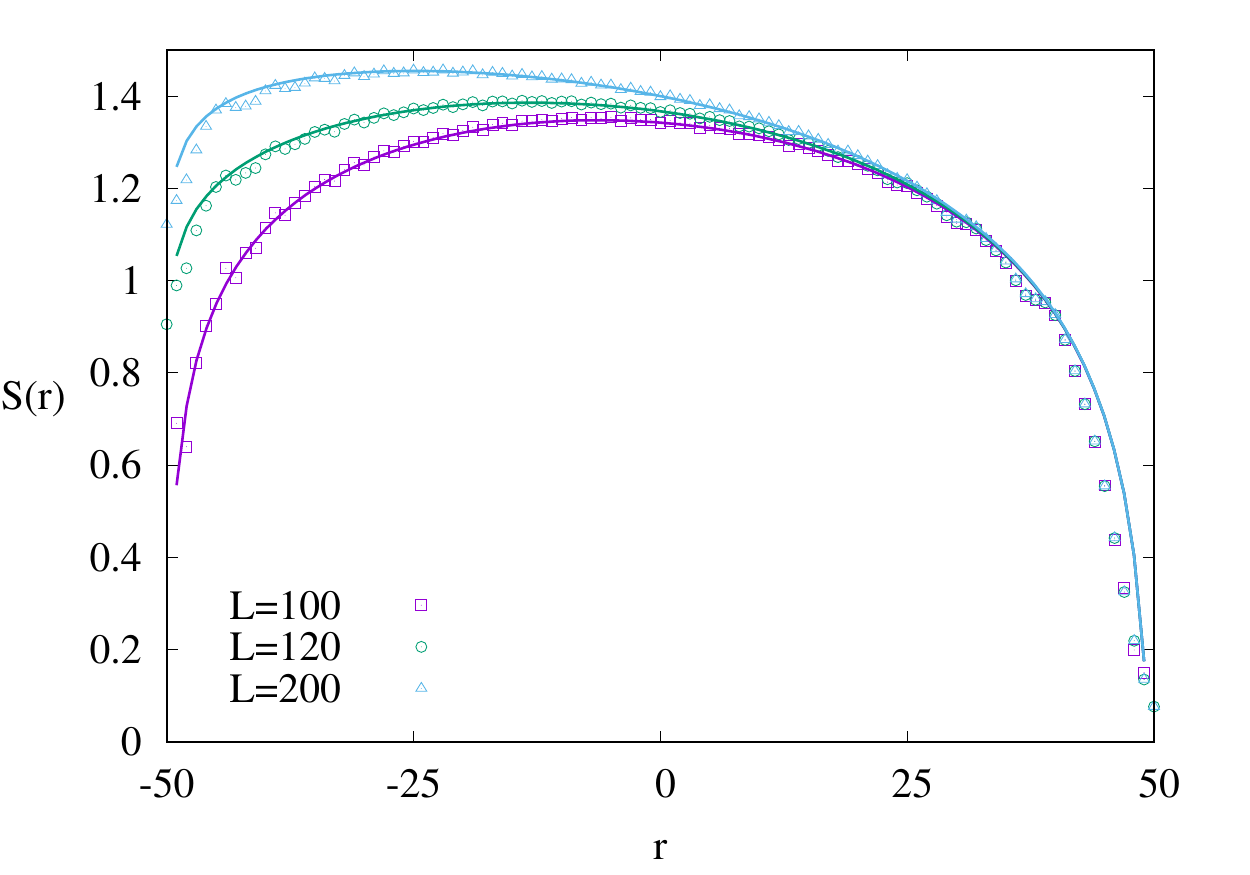}
\caption{Entropy profiles after the geometric quench for $t=50$ and various system sizes.
The solid lines correspond to the ansatz $S_g$ in Eq. \eqref{sg}. Only the front region $|r|<t$ is shown.}
\label{fig:sg}
\end{figure}

Finally, collecting the different contributions, one arrives at the result
%
\begin{align}
S_g &= \frac{1}{6}\ln \left[
\frac{L}{\pi} \sin\left(\frac{\pi (t-r)}{L}\right)\right] \nonumber \\
&+ \frac{1}{12}\ln \left[(t-r)\left(1-\left(r/t\right)^2\right) \right] + k,
\label{sg}
\end{align}
where $|r|<t$ and $k$ is a nonuniversal constant. In particular, setting
$r=0$ one recovers the ansatz put forward in \cite{AHM14}.
To test the result \eqref{sg}, we calculated the entropy profiles for free-fermion
chains using standard correlation matrix techniques \cite{PE09}.
Fig. \ref{fig:sg} shows the result for a fixed time $t=50$ and for various chain
sizes, compared to the ansatz \eqref{sg} shown by solid lines.
One sees a very good agreement with the numerical data.
The only free parameter is the constant, which was fixed at $k\approx 0.44$
by fitting the ansatz to one of the data sets. We also carried out
calculations for a larger time $t=100$ (not shown) with similarly good
agreement, confirming the validity of the result in Eq. \eqref{sg}.

%
%

\subsection{XXZ chain}

\begin{figure*}[t]
  \centering
  \noindent
  \includegraphics[trim={ 20px 40px 80px 0}]{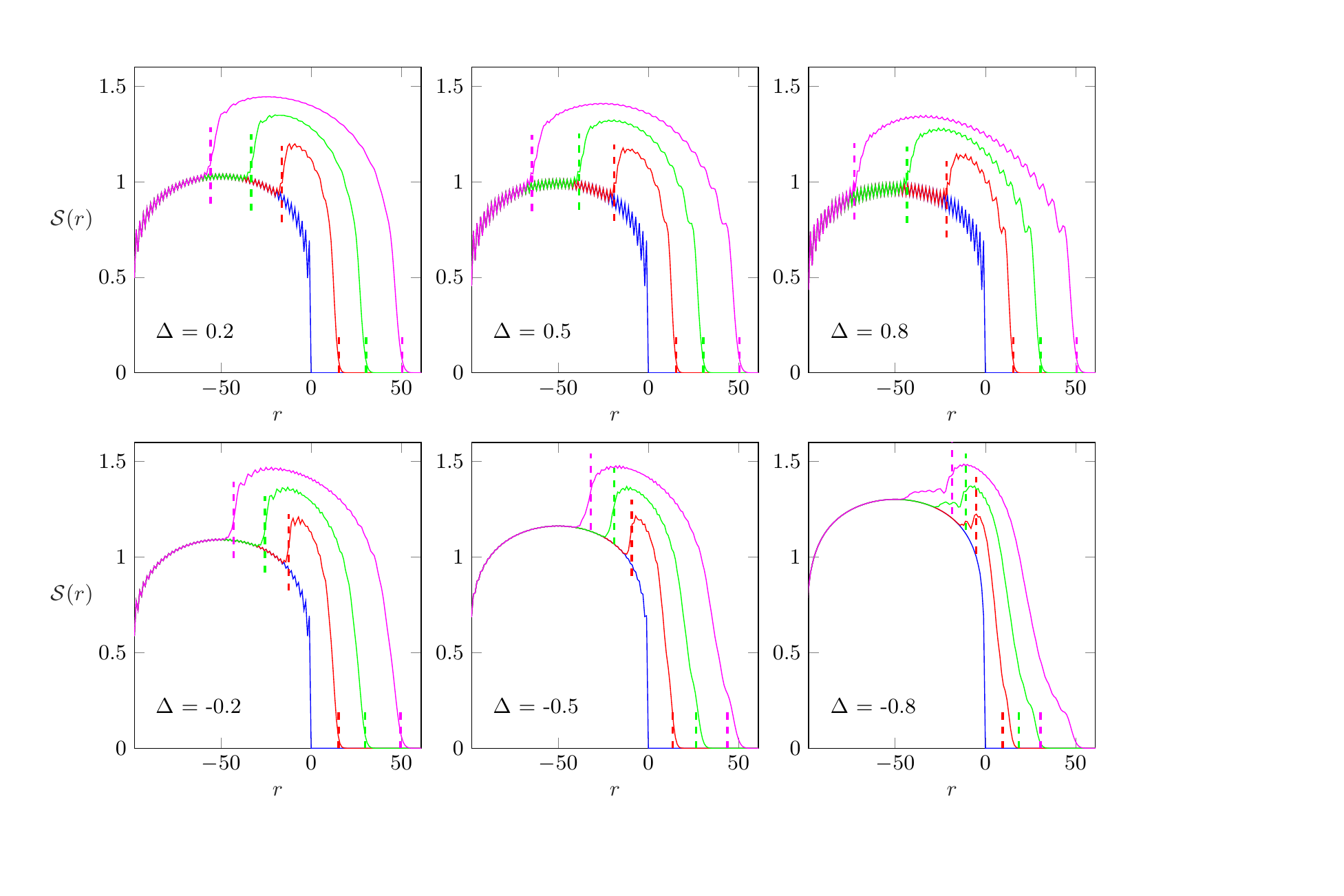}
  \caption{Entanglement profiles for different values of $\Delta$ at times
  $t = 0$ (blue), $t = 15$ (red), $t = 30$ (green) and $t = 50$ (magenta)
  after the quench. The GHD edges $r=\zeta_{min}t$ and $r=\zeta_{max}t$
  are marked by vertical dashed lines.}
      \label{fig:ent}
\end{figure*}

We continue with the numerical study of the entanglement profile for the XXZ chain.
In Fig. \ref{fig:ent} the results of DMRG calculations for a chain with $L = 200$ are shown.
The snapshots of the profiles are plotted for various times, and the $\Delta$
values considered are the same as for the magnetization in Fig. \ref{fig:sz}.
At $t=0$ (blue curve) the entanglement entropy is trivially vanishing for a cut across
the right half-chain, whereas the profile on the left is given by the well-known
CFT formula for the ground state \cite{CC04}. After the quench, the entanglement spreads
in both directions and a profile qualitatively similar to the XX case emerges.
However, one expects that the left and right edges of the front are given by $r=\zeta_{min} t$
and $r=\zeta_{max}t$, respectively, as indicated by the dashed lines in Fig. \ref{fig:ent}.
While for $\Delta > 0$ this seems to hold perfectly, for $\Delta<0$ one observes,
similarly to the magnetization profiles, a tail reaching beyond the GHD edges on both sides,
increasing for large negative values of $\Delta$.

\begin{figure}[H]
  \centering
  \noindent
  \includegraphics[]{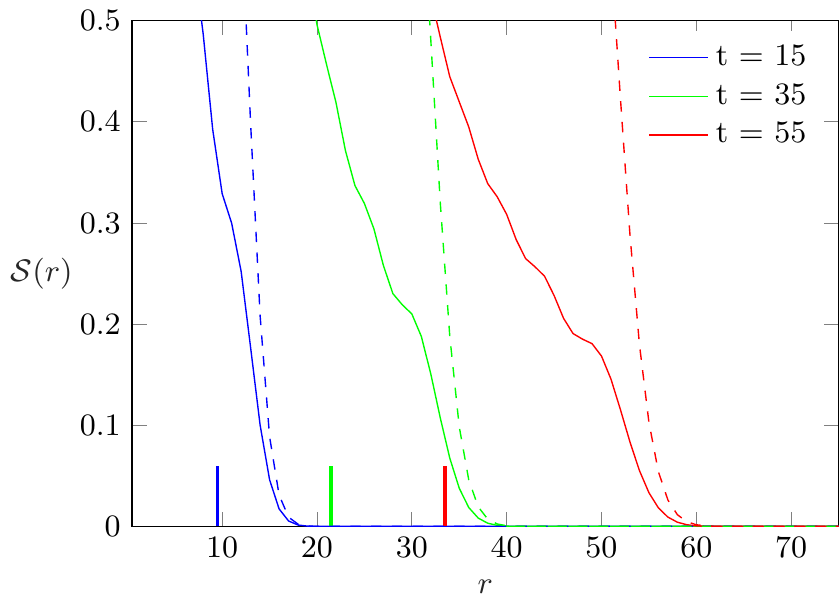} 
  \caption{Comparison of the right edge of the entanglement profile for
  $\Delta = -0.8$ (solid lines) vs. $\Delta = 0.8$ (dashed lines) and various times.
  The GHD edges $r=\zeta_{max}t$ for $\Delta = -0.8$ are indicated by vertical lines.} 
      \label{fig:entedge}
\end{figure}

It is instructive to have a closer look at the right tail of the front expanding into the vacuum.
As already discussed in the previous section, the tail behavior is reminiscent of the
domain-wall quench where, however, the dynamics is invariant under the change of sign
in $\Delta$. To emphasize the difference for the geometric quench, in Fig. \ref{fig:entedge}
we compare the edge entropy profiles between $\Delta = 0.8$ and $\Delta = -0.8$.
While in the repulsive case the profile has a sharp edge with an abrupt increase,
for the attractive one the free edge remains soft until reaching the GHD edge, where
the slope becomes steep. The profile between the soft and hard edges develops
a steplike structure, as can be seen for larger times in Fig. \ref{fig:entedge}.
In fact, beyond the left edge the profile develops a qualitatively similar tail,
which can already be seen on Fig. \ref{fig:ent} without magnifying the region.

Regarding the bulk profile, it is tempting to find a generalization to
the ansatz in \eqref{sg}. In fact, the CFT result \eqref{slocfin} for the local quench can be
applied to the XXZ case by explicitly including the spinon velocity, i.e. substituting
$t \to v_s t$, which we have verified by DMRG calculations.
On the other hand, however, the other constituent of the ansatz
originates from the domain-wall quench, where the result \eqref{sdw} is specific
to free fermions. Hence, despite the qualitatively similar behavior of the profiles,
the XXZ case can not simply be related to the XX result \eqref{sg} by rescaling
with the front velocities.

\section{Boundary Effects\label{sec:refl}}

So far we have only considered situations where the propagating
front does not reach the boundaries of the chain. Since the formulation
of GHD genuinely involves the thermodynamic limit, it is interesting to
ask what happens when finite size effects play a dominant role, i.e.
when reflections of the wavefront occur. 

\subsection{XX chain}

%
\begin{figure*}[htb]
\center
\includegraphics[width=\columnwidth]{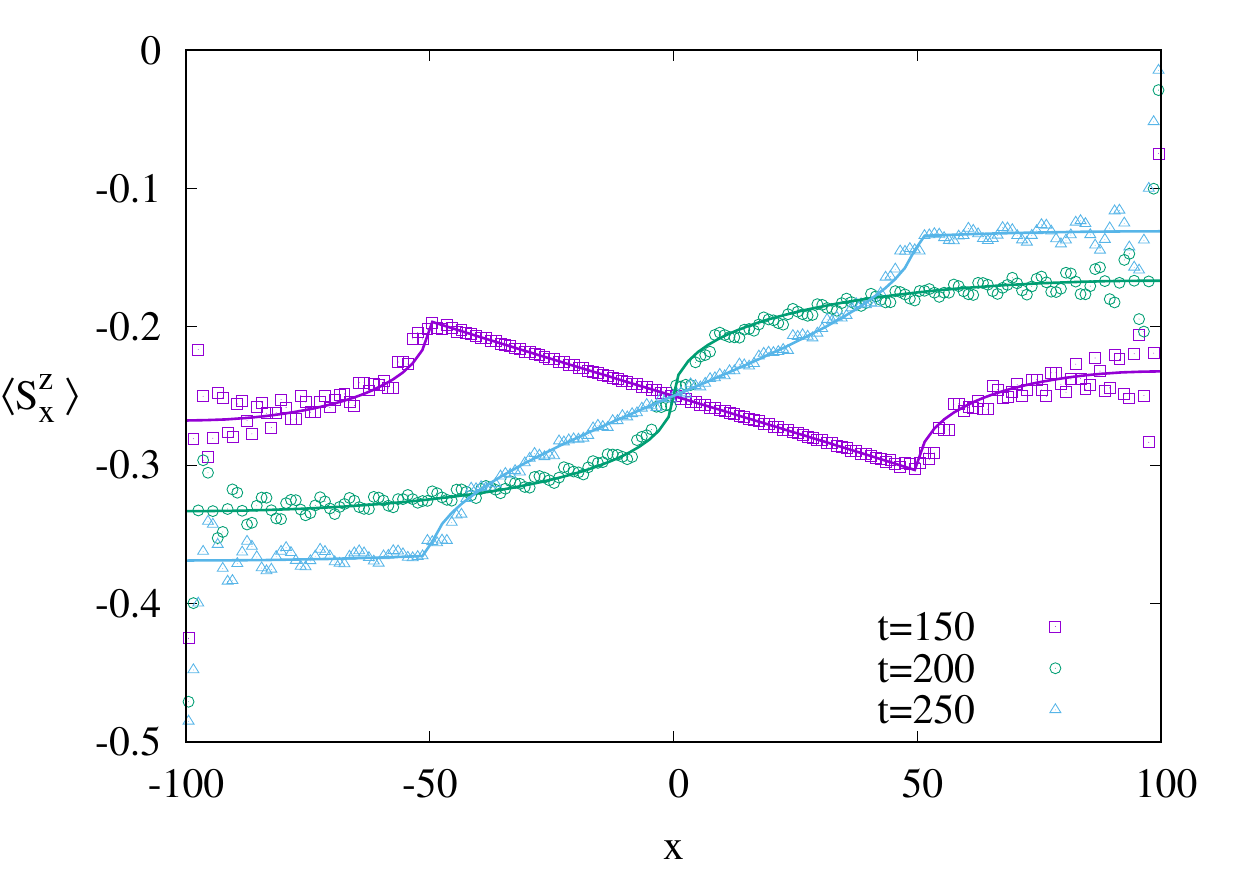}
\includegraphics[width=\columnwidth]{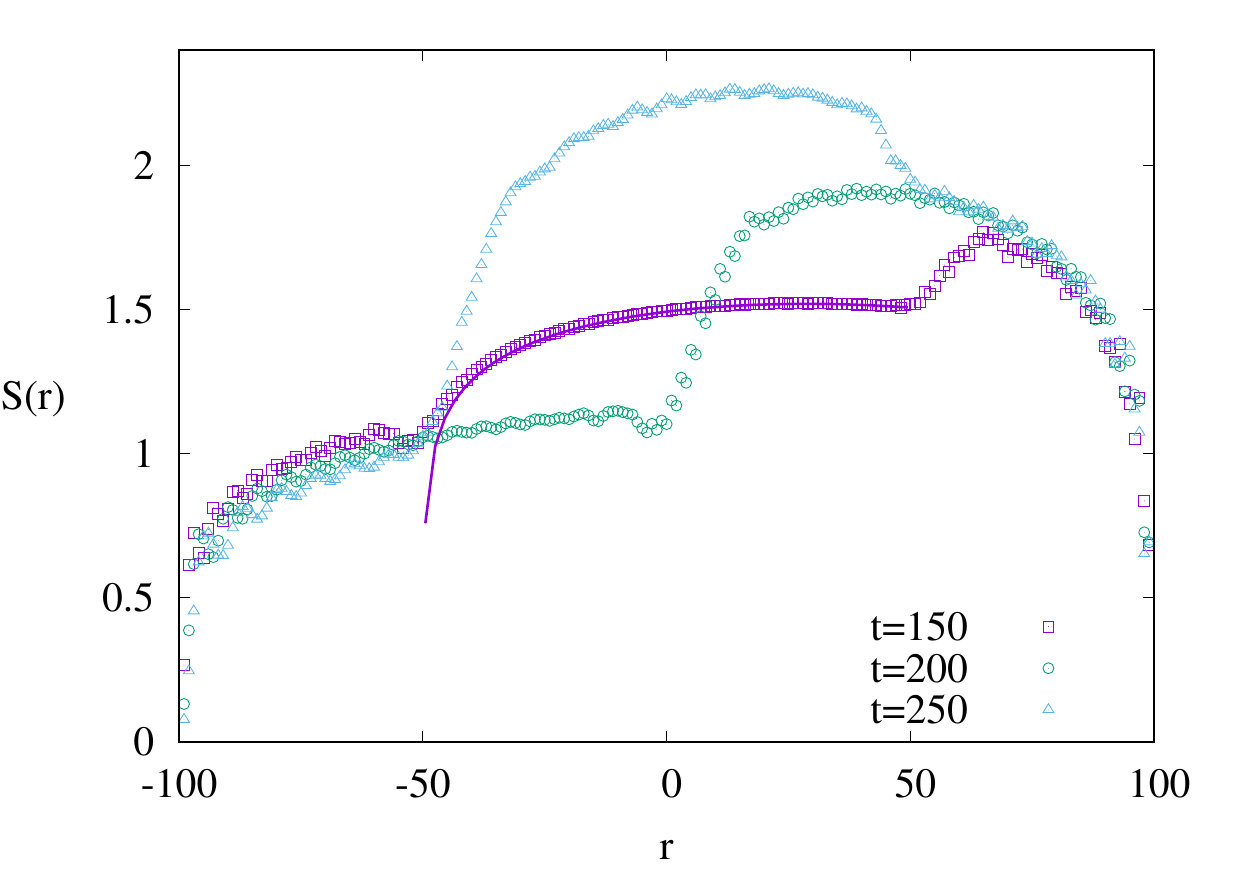}
\caption{Magnetization (left) and entanglement (right) profiles after reflection of the wavefront from the boundaries.
The solid lines show the results \eqref{szxx}, \eqref{szr1} and \eqref{szr2} for 
the magnetization and \eqref{sg} for the entropy, respectively.}
\label{fig:reflxx}
\end{figure*}
%

We start again by considering
the XX chain where, due to the complete independence of the quasiparticle
velocities from the mode occupations, the hydrodynamic picture remains applicable
even after reflections from the boundaries take place.
Indeed, determining the magnetization requires only a proper bookkeeping
of the contributions from the reflected particles. Considering
a fixed site with $x>0$ on the r.h.s. of the chain, the result \eqref{szxx}
remains true for times $t<L-x$, i.e. until the reflected particles with maximal
velocity $v_{max}=1$ arrive there. For larger times one simply adds the contribution
of the reflected density
\eq{
\braket{S^z_x} = -\frac{1}{2} + 
\mathcal{N} \left(\frac{x}{t} \right) +  \mathcal{N} \left(\frac{L-x}{t} \right),
\label{szr1}}
where $L-x<t<L+x$. This last requirement ensures, that only reflections
from the right end of the chain could take place.

For even larger times, one has to take into account the reflections
from the left boundary. To this end one should first note, that the left-moving
particles could be considered as \emph{holes} penetrating the originally
zero-magnetization background. This also follows directly from the exact
symmetry relation $\braket{S^z_{-x}}$=$-1/2-\braket{S^z_{x}}$, which
can be used to obtain the magnetization on the l.h.s. of the chain.
Hence, for times $t>L+x$, the contribution of the reflected holes should
appear as
\eq{
\braket{S^z_x} = -\frac{1}{2} + 
\mathcal{N} \left(\frac{x}{t} \right) +  \mathcal{N} \left(\frac{L-x}{t} \right)
- \mathcal{N} \left(\frac{L+x}{t} \right).
\label{szr2}}
The above result is then valid for times $L+x<t<2L-x$, i.e. until
the fastest holes arrive to site $x$ after a double reflection from
both left and right boundaries. Clearly, this pattern could be continued
to arbitrary times after multiple reflections, always adding the fermionic
density with the proper sign and argument.

The results \eqref{szxx}, \eqref{szr1} and \eqref{szr2} are compared to exact
numerical free-fermion calculations on the left of Fig. \ref{fig:reflxx}.
One observes that, apart from oscillations, the average
magnetization is well described by the semiclassical formulas.
The oscillations are rather strong around the boundaries and one expects
that, after many reflections, the profile becomes increasingly noisy.
On the right of Fig. \ref{fig:reflxx} we also plotted the corresponding entanglement
profiles. As one can see, the result in \eqref{sg} remains valid
for that part of the profile which is not yet reached by the reflected wavefront.
Interestingly, after each reflection one has a steady increase of
entanglement, which was already pointed out in \cite{AHM14} for $r=0$.
Unfortunately, however, a quantitative understanding of the profile is
still beyond our reach.

\subsection{XXZ chain}

\begin{figure*}[t]
  \centering
  \noindent
  \scalebox{0.8}{\includegraphics[trim={30px 33px 20px 20px}]{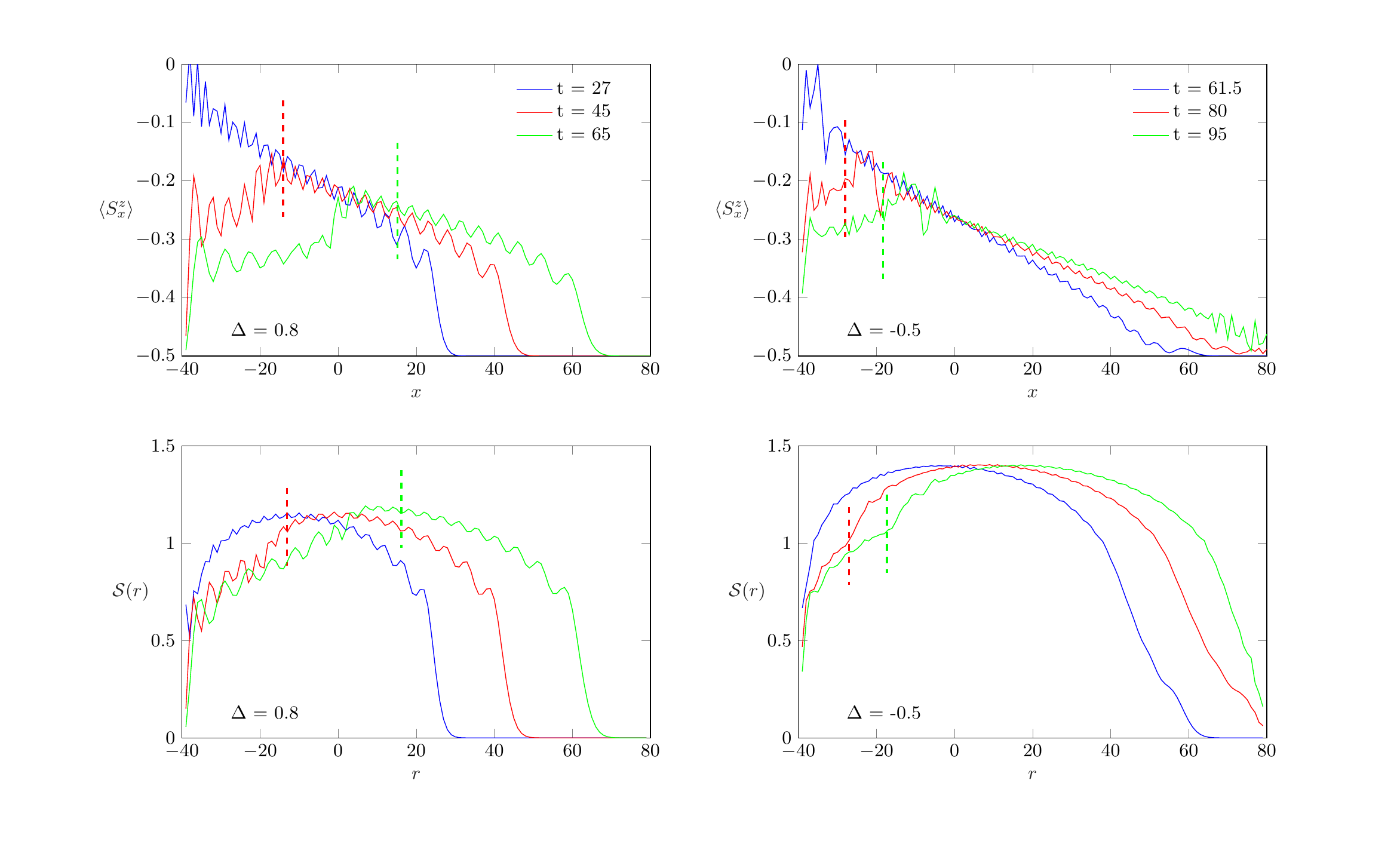}}
  \caption{Magnetization (top) and entanglement (bottom) profiles 
   for $\Delta = 0.8$ (left) and $\Delta = -0.5$ (right), just before (blue) and after the reflection
  (red, green) of the front from the left boundary of the chain. The data is plotted against the
  distance from the junction. The dashed lines indicate the edge positions corresponding to
  a reflected front with constant speed $v_s$.}
      \label{fig:reflxxz}
\end{figure*}

In contrast to the XX case, it is far from trivial how the hydrodynamic approach
could be extended to include reflected quasiparticles in the interacting case.
Here we try to understand only some simple qualitative features of the
dynamics after reflection, focusing on the front which propagates in the
l.h.s. of the system. In order to avoid interference with the reflection of the
right-propagating front, for this simulation we considered a chain of size
$L=L_1+L_2$, composed initially of two unequal pieces $L_1=40$ (ground state)
and $L_2=80$ (vacuum). 

Our results for both the magnetization and entropy profiles are shown in Fig.
\ref{fig:reflxxz} for two different anisotropies, with the colors corresponding
to different evolution times. The dashed lines indicate the calculated front
positions, assuming that the speed of propagation after reflection is still given
by the spinon velocity $v_s$. The blue curves correspond to times
$t=L_1/v_s$, i.e. when the front is just supposed to reach the boundary,
which is indeed what we observe in Fig. \ref{fig:reflxxz}. In contrast,
after reflection there is a clear mismatch between the calculated and
the actual edge locations: the front slows down for $\Delta>0$ and
speeds up for $\Delta<0$, the effect becoming more apparent for
larger times.
The change of speed is due to the fact, that
the reflected front does not any more propagate in a
zero-magnetization background, but rather in a nontrivial one left
behind by the primary front. Since this background is inhomogeneous,
we expect that the speed of the reflected front will actually change in time,
which is supported by our numerical data. 
A more detailed analysis is, however, difficult due to the ambiguity in
defining the edge of the reflected front, with its location getting
washed out by superimposed oscillations.

Regarding the entropy evolution, one should comment on the
previous observations made in Ref. \cite{AHM14}, where the following
ansatz for the entropy across the junction $r=0$ for times $t\ll 2L_2$
was put forward
\eq{
S(r=0)= \frac{1}{6}\ln \left[\sqrt{v_e t}
\frac{2L_1}{\pi} \sin\left(\frac{\pi v_e t}{2L_1}\right)\right] + \const.
\label{sr0}}
Note that this is nothing else but the XX result \eqref{sg} for $r=0$
and $L=2L_1$, after a rescaling $t \to v_e t$, where the parameter
$v_e$ was interpreted as an entanglement spreading rate.
Indeed, $t =2L_1/v_e$ should correspond to the roundtrip time of the
entanglement front and the speed $v_e$ was obtained by fitting
the ansatz \eqref{sr0} to the data,
with the result $v_e < v_s$ for $\Delta=0.5$ and $v_e \gtrapprox v_s$
for $\Delta=-0.5$ (see Fig. 12 of Ref. \cite{AHM14}). This is in
perfect accordance with our observations in Fig.~\ref{fig:reflxxz}.
However, instead of being an entanglement spreading rate,
the correct interpretation of $v_e$ is due to the modified quasiparticle
velocity in the inhomogeneous background. Indeed, the very same
effect appears also in the magnetization profile. Remarkably, even though the
front velocity appears to be time dependent after reflection, the simple
ansatz \eqref{sr0} was found to give a rather good description of the
entropy for $t < 2L_1/v_e$, with $v_e$ being the average roundtrip velocity.

\section{Fluctuations vs. entropy\label{sec:fluct}}

\begin{figure*}[t]
  \centering
  \noindent
  \includegraphics[trim={9px 38px 0 0}]{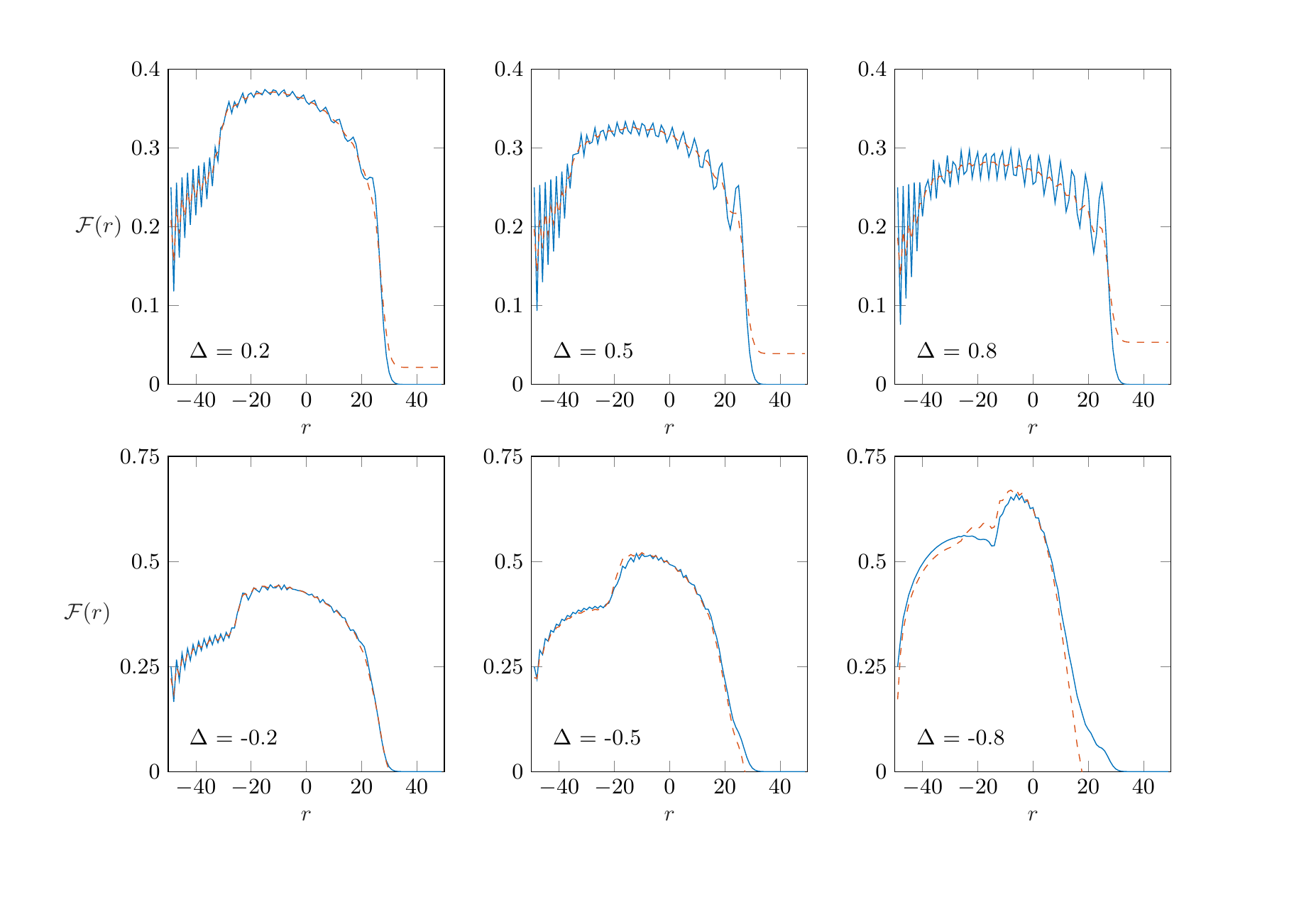}
  \caption{Comparison between the magnetization fluctuations $\mathcal{F}(r)$ (blue) and the
  scaled entanglement entropy $S(r)$ (red), according to Eq. \eqref{FS}, for a system of $L=100$
  and at $t=30$.}
      \label{fig:FS}
\end{figure*}

To conclude our studies of the geometric quench, we shall consider yet another
physical quantity, namely the profile of the magnetization fluctuations.
Since the XXZ dynamics conserves the overall magnetization, the fluctuations
are clearly vanishing for the full chain. However, considering only a segment $A$
(see Fig. \ref{fig:setup}), the subsystem fluctuations can be defined as
\begin{equation}
  \mathcal{F} = \expval{\left( \sum_{i \in A} S^z_i - \expval{\sum_{i \in A} S^z_i} \right)^2},
  \label{F1}
\end{equation}
where the expectation values are taken with respect to the time evolved state \eqref{psit}.
Note that, in the fermion language, $\mathcal{F}$ is equivalent to the variance of the
particle number in $A$.

For free fermion systems, the study of fluctuations is motivated by an exact relation
between the ground-state entanglement entropy and the particle number statistics \cite{KL09},
reproducing the entropy as a cumulant series \cite{Song11,CMV12}.
The scaling of the variance has thus been extensively studied in the ground state
of the XX chain \cite{ELR06,Song12} as well as out of equilibrium for the simple domain-wall
initial state \cite{Schoenhammer07,AKR08}. In all of the above mentioned cases
one finds that, to leading order, the entropy is simply proportional to the variance,
whereas the higher order cumulants give only subleading contributions.

Although the cumulant series relation between entropy and fluctuations roots deeply
in the free-fermion nature of the state, there are some known
extensions to interacting systems. In particular, for critical ground states described by
a Luttinger liquid, the fluctuations were also found to be proportional to the entropy \cite{Song10}
\begin{equation}
  \mathcal{F} \simeq K\frac{3}{\pi^2} S + \const .
  \label{FS}
\end{equation}
Here $K$ denotes the Luttinger parameter, while the constant is non-universal.
The relation \eqref{FS} has been checked explicitly for the XXZ ground state \cite{Song10},
where the Luttinger parameter is known from the Bethe ansatz solution
\begin{equation}
  K = \frac{1}{2} \left( 1 - \frac{\acos(\Delta)}{\pi} \right)^{-1}.
\label{K}
\end{equation}
However, to the best of our knowledge, no such relation has been established
in an out of equlibrium context so far.

Our goal here is to study the fluctuations after the geometric quench,
which can also be rewritten as a sum over correlation functions
\begin{equation}
  \mathcal{F} = \sum_{i,j \in A} \left[ \expval{S_i^z S_j^z} - \expval{S_i^z} \expval{S_j^z} \right] .
  \label{F2}
\end{equation}
Although these objects are straightforward to evaluate via DMRG, one needs the full
matrix of correlators within the subsystem. This makes the computation somewhat more
demanding, thus the simulations are now performed on a smaller chain with $L = 100$ sites.
The fluctuation profile $\mathcal{F}(r)$ is measured at time $t = 30$, and is shown by the 
blue lines in Fig. \ref{fig:FS} for a set of interaction parameters $\Delta$.
The front region is clearly visible and qualitatively similar to the entropy profiles.

In order to test the relation \eqref{FS} between entropy and fluctuations, we have fitted the
constant for the region of the profile which corresponds to the ground state (i.e. outside the lightcone).
This was done by first smoothening out the oscillations in the data and then minimizing the difference
between the corresponding profiles $\mathcal{F}(r)$ and $\mathcal{S}(r)$.
With the fitted constant, one can now compare the profiles in the entire front region
by plotting the ansatz \eqref{FS}, shown by the red dashed lines in Fig. \ref{fig:FS},
together with $\mathcal{F}(r)$. Quite remarkably, the two profiles show a good
agreement also within the front region, up to the superimposed oscillations. 
The collapse is particularly good for moderate values of $\Delta$, while for larger
negative values the curves start to differ increasingly (for large $\Delta >0$ the
oscillations dominate the profile and the comparison is hard).

The fact that Eq. \eqref{FS} seems to give a decent approximation also in
the far-from-equilibrium front region is rather intriguing, since the Luttinger parameter $K$
in Eq. \eqref{K} is calculated for the ground state. To have a better understanding of
this result, one should try to analyze the behavior of correlation functions in \eqref{F2},
which we leave for further studies.

\section{Conclusions\label{sec:conc}}

We have investigated the time evolution after a geometric quench in the XXZ chain and
showed that the magnetization profiles are nicely captured by generalized hydrodynamics.
While the entanglement profile is harder to describe within the hydrodynamic picture,
we were able to put forward an ansatz for the noninteracting case which shows a very good
agreement with the DMRG data. 

In order to arrive at our ansatz \eqref{sg}, we had to apply some heuristic arguments,
expressing the entropy production in the geometric quench as a kind of mixture of
local and domain-wall quenches. It would be desirable to put this result
on a firm ground, e.g. by a direct CFT treatment along the lines of
Ref. \cite{ADSV16}, identifying the curved-space metric corresponding to the
inhomogeneous time-evolved state. This might also allow for a generalization
to initial states with arbitrary fillings on both sides. Ideally, however, one would like
to cast the entropy as a sum over contributions from the different quasimomenta,
analogously to what has been found for global quenches \cite{AC17},
which would enable us to solve the interacting problem as well.
Whether such a representation is possible in situations with a logarithmic
entropy growth is still unclear.

Another interesting aspect is the physics of the edge, which was shown
\cite{ER13,VSDH15,PG17,Kormos17,Fagotti17}
to display a universal Tracy-Widom scaling \cite{TW94} for free fermions.
Clearly, the situation is more complicated in the interacting case, since
one has a splitting between the GHD edge and the free edge. Recent studies
for the domain-wall quench hint towards the possibility that the free tail
is characterized by a Tracy-Widom-like $t^{1/3}$ length scale \cite{BK18},
while the GHD edge seems to spread diffusively as $t^{1/2}$ \cite{Stephan19}.
We believe that the vacuum edge of the geometric quench may belong to the same
type of edge universality as observed for the domain wall. Additionally, however, one
has another edge appearing in our problem which connects to the ground-state region
and might display a different type of behavior. A detailed study of these edge
phenomena requires much more numerical effort and is left for future studies.

Finally, it would be illuminating to understand how the presence of boundaries
could be reconciled with the theory of generalized hydrodynamics. One feature
we observed is that the edge velocity becomes time dependent after reflection,
due to propagation in a nontrivial inhomogeneous background. Whether
a quantitative description of the reflected front is possible along the lines of
GHD is an interesting question to be addressed.

\begin{acknowledgments}

We thank J. Viti and J.-M. St\'ephan for a fruitful discussion.
The authors acknowledge funding from the Austrian Science Fund (FWF) through
project No. P30616-N36.

\end{acknowledgments}


\onecolumngrid

\appendix*
\section{Perturbative calculation of the edge around $\zeta_{max}$}
\label{sec: app1}

As discussed in the main text, for $\Delta>0$ the GHD solution around
the rightmost ray $\zeta_{max}=1$ is given by the occupation function
\eqref{nzeta2} by solving \eqref{vl12}. We assume that, sufficiently close to the GHD edge,
the interval $\left[\lambda_1,\lambda_2\right]$ of occupied rapidities remains small,
i.e. $|\lambda_{1,2}-\tilde\lambda| \ll 1$ with $\tilde\lambda$ given by \eqref{lamt}.
For such an occupation, the dressing of a function $f$ can be considered as a perturbation
around its bare value
\begin{equation}
    f^{dr}(\lambda) \approx f(\lambda) + \delta f(\lambda) \, .
    \label{eq: fdr}
\end{equation}
\noindent
Inserting into the dressing equation \eqref{eq: dress} one obtains
\begin{equation}
  f(\lambda) + \delta f(\lambda) + 
  \int_{\lambda_1}^{\lambda_2}  \frac{\dd \mu}{2 \pi} \mathcal{K}(\lambda - \mu)
  \left[ f(\mu) + \delta f(\mu) \right] = f(\lambda) \, .
\end{equation}
To extract the leading order in the perturbation series, we neglect the term
$\delta f$ within the integral and expand all the functions to first order in $\nu=\lambda-\mu$,
which leaves us with
\begin{equation}
 \delta f(\lambda) +
  \int_{\lambda-\lambda_2}^{\lambda-\lambda_1} \frac{\dd \, \nu}{2 \pi}
  \left(\mathcal{K}(0) + \nu \mathcal{K}'(0)\right)
  \left( f(\lambda) - \nu f'(\lambda) \right) = 0 \, .
\end{equation}
Setting $\lambda_1=\tilde \lambda - \epsilon_1$, $\lambda_2=\tilde \lambda + \epsilon_2$ and
carrying out the integrals, we finally arrive at
\begin{equation}
2\pi \, \delta f(\lambda) =
 -\mathcal{K}(0) f(\lambda)(\epsilon_1+\epsilon_2)
 -\left[ \mathcal{K}'(0)f(\lambda) - \mathcal{K}(0) f'(\lambda)\right]
 \frac{\epsilon_1^2 - \epsilon_2^2+2(\lambda-\tilde\lambda)(\epsilon_1+\epsilon_2)}{2}
 + \mathcal{O}(\epsilon^3) \, .
   \label{deltaf}
\end{equation}

With this result at hand, we can now calculate from Eq. \eqref{vdr} the dressed velocity
\begin{equation}
  v \approx \frac{e' + \delta e'}{p' + \delta p'} \approx 
  \frac{e'}{p'}\left[ 1 + \frac{\delta e'}{e'} - \frac{\delta p'}{p'} + 
  \left(\frac{\delta p'}{p'}\right)^2 - \left(\frac{\delta e'}{e}\right)\left(\frac{\delta p'}{p'}\right)\right] ,
\end{equation}
where we have droppped the arguments $\lambda$.
Applying \eqref{deltaf} for both $\delta e'$ and $\delta p'$ and keeping only up to quadratic terms
in $\epsilon_{1,2}$, one has
\eq{
v \approx v_0
\left[ 1 + \frac{\mathcal{K}(0)}{2\pi} \left(\frac{e''}{e'} - \frac{p''}{p'}\right)
\frac{\epsilon_1^2 - \epsilon_2^2+2(\lambda-\tilde\lambda)(\epsilon_1+\epsilon_2)}{2} \right],
}
where $v_0$ is the bare velocity, see Eq. \eqref{v0}.
The correction to the bare velocity depends on the ratios of second and first
derivatives of the energy and momentum. Since the factor multiplying them is already
quadratic in $\epsilon_{1,2}$, it is enough to evaluate the ratios at $\lambda = \tilde {\lambda}$.
Interestingly, however, a simple calculation leads to the result
\begin{equation}
  \eval{\frac{e''}{e'}}_{\lambda = \tilde{\lambda}} =  \eval{\frac{p''}{p'}}_{\lambda = \tilde{\lambda}}.
\end{equation}
\noindent

Hence, to leading order, the velocity around the edge is just given by its bare value,
with corrections $\mathcal{O}(\epsilon^3)$. The rapidities $\lambda_{1,2}$ then
follow from the condition $v_0(\lambda_1)=v_0(\lambda_2)=\zeta$. Expanding
the bare velocity around $\tilde \lambda$ gives
\eq{
v_0(\tilde\lambda\pm\epsilon) \approx v_0(\tilde\lambda) 
\pm \epsilon v'_0(\tilde\lambda) + \frac{\epsilon^2}{2} v''_0(\tilde\lambda) \, .
}
However, as discussed in the main text, $\tilde\lambda$ is exactly the maximum
of the bare velocity, $v'_0(\tilde\lambda)=0$, with its value given by $v_0(\tilde\lambda)=1$.
Furthermore, the second derivative can be calculated as $v''_0(\tilde\lambda)=-\cot^2(\gamma)$
and thus we get
\begin{equation}
  v_0(\tilde{\lambda} \pm \epsilon) = 1 - \frac{\epsilon^2}{2} \cot^2(\gamma) = \zeta \, .
\end{equation}
Solving for $\epsilon$ then leads to the result
\begin{equation}
  \epsilon = \sqrt{2 (1-\zeta)} \tan(\gamma) \, ,
  \label{eps}
\end{equation}
reported in Eq. \eqref{epsz} of the main text.

It should be stressed that, for the perturbation theory to work, the condition
$\epsilon \ll 1$ must be satisfied. From \eqref{eps} one can see that this becomes
problematic, as the interaction strength is decreased. Indeed, for $\Delta \to 0$
($\gamma \to \pi/2$) one has $\epsilon \to \infty$, i.e. the solution diverges.
The reason is that for very small interactions, the bare velocity develops only
a very tiny maximum around extremely high rapidities $\tilde\lambda \gg 1$,
thus deteriorating the quality of the approximation.
In particular, at the free-fermion point $\Delta=0$ the approximation
fails completely, which can also be seen by expanding the analytical result \eqref{szxx}
for the magnetization profile around $\zeta=1$. This gives to leading order
\eq{
\braket{S^z} \approx -\frac{1}{2} + \frac{1}{2\pi}\sqrt{2(1-\zeta)},
\label{szxxedge}}
where the coefficient of the square-root is off by a factor of $2$ with respect to the
approximation \eqref{szedge}. Moreover, even considering a larger value
$\Delta=0.8$ and requiring $\epsilon < 0.1$, one has from \eqref{eps} the condition
$\zeta > 0.99$. This explains why the approximate profile deviates essentially
immediately from the GHD solution in Fig. \ref{fig:sz}.

\begin{figure}[H]
  \centering
  \includegraphics[]{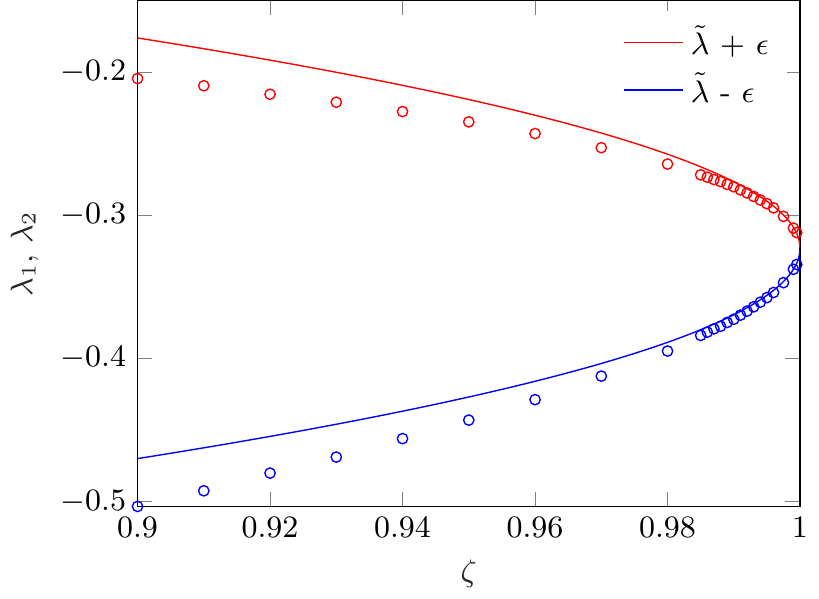}
  \includegraphics[]{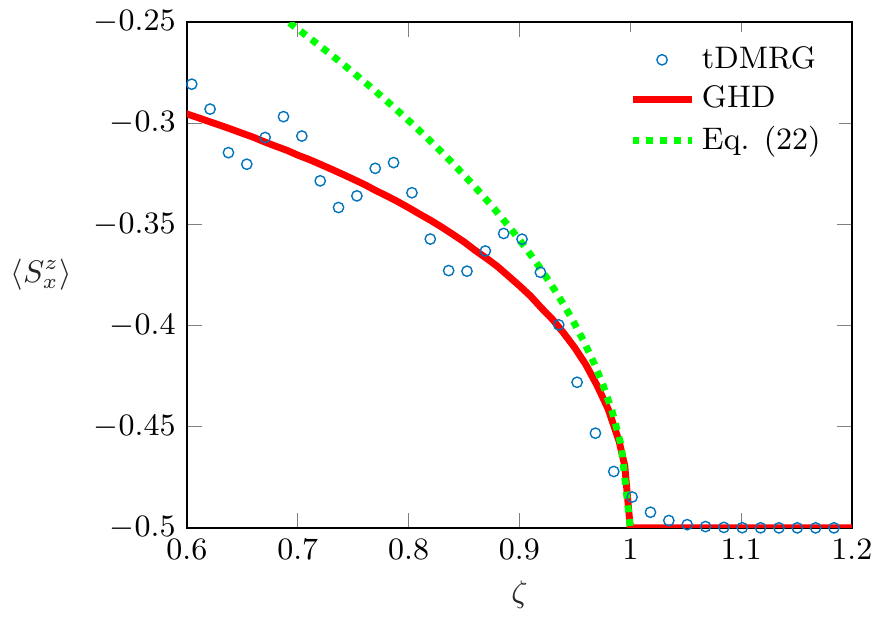}
  \caption{Left: rapidities $\lambda_1$ (blue circles) and $\lambda_2$ (red circles)
  as obtained from the iterative numerical solution of Eq. \eqref{vl12} for $\Delta=0.95$.
  The approximate solutions are shown by solid lines, see Eq.~\eqref{eps}, with a good
  agreement near $\zeta = 1$. Right: corresponding edge magnetization profile and 
  its approximation.}   
      \label{fig:app}
\end{figure}

On the other hand, one expects that the edge approximation should perform
much better for $\Delta \to 1$, i.e. around the isotropic Heisenberg point.
This is indeed what we observe by comparing it to the numerical solutions
$\lambda_1$ and $\lambda_2$ of the GHD ansatz \eqref{vl12}.
This is shown in the left of Fig. \ref{fig:app} for $\Delta=0.95$, where
the approximation appears to be rather good around the edge
but deviates as one moves further away. On the right of Fig. \ref{fig:app}
we also show the edge profile for the magnetization, comparing
the GHD solution to the approximation \eqref{szedge}
and to the tDMRG data for $L=200$ and $t=60$.

\twocolumngrid

\bibliographystyle{apsrev.bst}
\bibliography{literatur}

\end{document}